\documentclass[preprint,aps,12pt,preprintnumbers,eqsecnum,nofootinbib]{revtex4}
\usepackage{graphicx}

\usepackage{color}
\usepackage{amssymb,amsmath,amsfonts}
\usepackage{epstopdf} 
\usepackage{braket}

\newcommand{\be}{\begin{equation}}
\newcommand{\ee}{\end{equation}}
\newcommand{\bea}{\begin{eqnarray}}
\newcommand{\eea}{\end{eqnarray}}

\newcommand{\Pl}{\text{P}}

\begin{document}

%
%
\title{
Composite graviton self-interactions \\  in a model of emergent gravity
\vskip 0.1in}
\author{Christopher D. Carone}\email[]{cdcaro@wm.edu}\affiliation{High Energy Theory Group, Department of Physics,
College of William and Mary, Williamsburg, VA 23187-8795, USA}
\author{Tangereen V. B. Claringbold}\email[]{tvclaringbold@email.wm.edu}\affiliation{High Energy Theory Group, Department of 
Physics, College of William and Mary, Williamsburg, VA 23187-8795, USA}
\author{Diana Vaman}\email[]{dv3h@virginia.edu}\affiliation{Department of Physics, University of Virginia,
Box 400714, Charlottesville, VA 22904, USA}
\date{\today}
\begin{abstract}
We consider a theory of scalars minimally coupled to an auxiliary background metric.  The theory is generally covariant and subject to the 
constraint of vanishing energy-momentum tensor.  Eliminating the auxiliary metric leads to a reparametrization invariant, non-polynomial, 
metric-independent action for the scalar fields. Working in the limit of a large number of physical scalars, a composite massless spin-$2$ state, the graviton, 
was identified in previous work, in a two-into-two scalar scattering process.  Here, we further explore the possibility that dynamical emergent gravity is a natural feature 
of generally covariant quantum field theories, by studying the self-interactions of the emergent composite graviton.  We show that  the fine-tuning previously imposed 
to ensure the vanishing of the cosmological constant, as well as the existence of the massless spin-$2$ state, also assures that the emergent graviton's cubic 
self-interactions are consistent with those of Einstein's general relativity, up to higher-derivative corrections. We also demonstrate in a theory with more than one type of 
scalar that the composite graviton coupling is universal.  
\end{abstract}
\maketitle


\section{Introduction}

In previous work~\cite{Carone:2016tup},  Carone, Erlich and Vaman presented a theory of $D+N$ scalar fields in which the graviton emerges 
as a massless composite spin-$2$ bound state.   A number of ingredients were essential as organizing principles in the construction of the model.  
First, the tree-level action, which included a non-propagating, auxiliary metric field, was of a form that preserves general covariance.  
There would be little hope of recovering a massless composite graviton without such a symmetry requirement.  For the same reason, the 
regulator for divergent loop diagrams was required to preserve this symmetry as well.  Second, the energy-momentum tensor of the theory was 
exactly vanishing.  Aside from aesthetic simplicity, this condition assures that the emergence of a massless, composite graviton is not in conflict with the 
Weinberg-Witten theorem~\cite{Weinberg:1980kq}.  It also allows one to solve for the auxiliary metric field and eliminate it from the theory, leading to a 
non-polynomial and metric-independent form for the action.   When the coordinate invariance of the theory is gauge fixed, $D$ of the scalars, called 
clock-and-ruler fields in Ref.~\cite{Carone:2016tup}, are gauged away, so that the graviton couples to the non-vanishing energy-momentum tensor of 
the remaining, physical degrees of freedom.  The main result of Ref.~\cite{Carone:2016tup} followed from the calculation of a two-into-two scattering 
amplitude of scalars, to all orders in perturbation theory and at leading order in $1/N$, where $N \gg 1$ refers to the physical scalar degrees of freedom.   
The scattering amplitude was shown to contain a massless pole with an associated Lorentz-index structure consistent with the gauge-invariant part
of the graviton propagator. We review the model of Ref.~\cite{Carone:2016tup} in Section~\ref{sec:model}.

The idea that gravity may be the consequence of quantum corrections in a generally covariant non-gravitational theory is not new.   An early idea 
due to Sakharov~\cite{sakharov} is induced gravity: In this approach, one begins with a quantum field theory (QFT) in a classical 
background geometry.  Upon integrating out the QFT degrees of freedom, the one-loop partition function will contain the Einstein-Hilbert term for the 
background metric, a cosmological constant, and higher-order curvature terms.  The metric is not quantized, so induced gravity is a semi-classical 
theory~\cite{visser}.  In contrast, the starting point of our emergent gravity scenario is a non-metric QFT\footnote{We would like to point out that though we use the 
term ``emergent gravity", this is not the same as the subject of many recent works in which gravity is understood as emerging via entanglement in an underlying 
microscopic theory (see for example \cite{verlinde}).}.  In Ref.~\cite{Carone:2016tup}, it was shown how the sum over an infinite number of loops in this scalar 
theory leads to a pole associated with the exchange of a massless spin-$2$ particle, leading to the identification of the graviton as a composite state.  In this 
paper, we go beyond the linearized order to include the self-coupling of the emergent gravitons. In past models of composite gravitons, going beyond linearized 
gravity was a notable stumbling block~\cite{ohanian,sindoni}. For other related work on gravity as an emergent phenomenon in theories of matter, see 
Ref.~\cite{many}.

In the present work, we consider a number of issues that were not resolved in Ref.~\cite{Carone:2016tup}.  The two-into-two scattering calculation of Ref.~\cite{Carone:2016tup} yielded 
information on the emergent graviton two-point function, but not on gravitational self interactions.  Here we study a scattering amplitude involving six external scalar lines to extract useful information
on the emergent three-graviton vertex.  We show that the scattering amplitude has the correct form in the limit where it can be compared to the expectations 
of the weak field expansion of general relativity.  As noted in the literature review of Ref.~\cite{sindoni}, previous composite gravity models have succeeded in achieving
a massless spin-$2$ pole, but have failed to produce the correct graviton self-interactions.  Aside from providing a crucial consistency check of Ref.~\cite{Carone:2016tup},
our results represent distinct progress beyond past attempts at constructing consistent models with composite gravitons.

In both the scattering calculations of Ref.~\cite{Carone:2016tup} and the present work, scalar loops are 
regulated by dimensional regularization, with a finite regularization parameter $1/\epsilon$, which we take as a place holder for whatever generally covariant  
physical regulator may follow from a realistic ultraviolet completion of the theory.  As a consequence, the reduced Planck scale $M_\Pl$ was identified in 
Ref.~\cite{Carone:2016tup} as a function of $1/\epsilon$. The same approach is applied in our study of the emergent three-graviton vertex and we show that 
the identification of $M_\Pl$ in both calculations are in agreement.    We also clarify a technical point that was not explained in Ref.~\cite{Carone:2016tup}:  
our calculations follow from a perturbative expansion about a gauge-fixed background configuration for the clock-and-ruler fields.  While this implies that our 
results should maintain a form consistent with the general covariance of the action, it is not clear how exactly this comes about. We show that in our perturbative 
approach this is due to a cancellation of tree-level and loop effects that is manifest in our study of the three-graviton vertex.

Finally, we touch on two other variations of the original scenario.  In the first, we consider a simple extension of the theory in which there are two distinct sets of 
physical scalars with differing mass, to test the universality of the graviton couplings.  We show that in this case that the graviton pole persists, and that the 
graviton couples to the energy-momentum tensor for each set of scalars with a common Planck mass that depends on all the parameters of the theory.   In the 
second, we show how the effects of background matter can be taken into account by coupling the composite graviton operator to a classical background source, 
so that the graviton acquires a non-zero vacuum expectation value.   Duff~\cite{duff} showed in the context of general relativity how the presence of the 
background source modifies the flat-space metric in a way that is consistent with a desired curved background, and demonstrated this by reproducing the 
mass-expansion of the Schwarzschild metric.   We identify the Feynman diagrammatic expansion in the present model that reproduces the expectation value for the graviton that is relevant in this approach.

Our paper is organized as follows.  In the next section, we summarize the model of Ref.~\cite{Carone:2016tup}, and correct a number of sign errors that are important in 
understanding how general covariance is maintained in our perturbative approach.  In Sec.~\ref{sec:contribs}, we isolate the scalar interactions that are relevant for generating the 
emergent three-graviton coupling via loop effects.  In Sec.~\ref{sec:3gmp}, the loop calculations are performed and the results are compared with the expectations of the weak-field 
expansion of general relativity.  In Sec.~\ref{sec:uni}, we consider the extension of the theory to more than one set of distinct scalars and demonstrate the universality of the graviton couplings.  
In Sec.~\ref{sec:dff}, we discuss one approach to incorporating curved background into the theory.  In Sec.~\ref{sec:conc}, we summarize our conclusions.

\section{The model, with signs corrected} \label{sec:model}

Our starting point is the same as that of Ref.~\cite{Carone:2016tup}. For the sake of completeness we recall here the main features of the theory we are studying. We are also using this opportunity to correct some sign errors in Ref.~\cite{Carone:2016tup}.  We begin by considering a theory of of $N+D$ scalar fields in a curved $D$-dimensional spacetime described by a background metric $g_{\mu\nu}$, where $\mu$, $\nu\in\{0,1,\dots,D-1\}$:
\begin{equation}
S=\int d^Dx\,\sqrt{|g|}\left[\frac{1}{2}g^{\mu\nu}\left(\sum_{a=1}^N\partial_\mu\phi^a\,\partial_\nu\phi^a+\sum_{I,J=0}^{D-1}\partial_\mu X^I\partial_\nu X^J\eta_{IJ}\right)
-V(\phi^a)\right], \label{eq:S1}
\end{equation}
where $g=\det(g_{\mu\nu})$, $g^{\mu\nu}$ is the inverse of the metric $g_{\mu\nu}$, and $\eta_{IJ}$ are constants which we take to have the values of the Minkowski metric in $D$ 
dimensions. (We use the mostly-minus convention for the signature of $\eta_{IJ}$ and the metric $g_{\mu\nu}$.)  The defining feature of the action in Eq.~(\ref{eq:S1}) is that it is invariant under general coordinate transformations, with the fields $X^I$ and $ \phi^a$ transforming as scalars and the background field $g_{\mu\nu}$ transforming as a metric tensor. The background metric $g_{\mu\nu}$ has no dynamics, {\em i.e.}, there is no Einstein-Hilbert term in the action.   The theory is defined via functional integration over field configurations subject to the constraint of
vanishing energy-momentum tensor; the partition function may be written
\begin{equation}
Z=\int_{T_{\mu\nu}=0} {\cal D}g_{\mu\nu}\, {\cal D}\phi^a\, {\cal D}X^I\,e^{iS[\phi^a,\, X^I, \,g_{\mu\nu}]} \,\,\, ,
\end{equation}
where the energy-momentum tensor is given by
\bea
T_{\mu\nu}(x)&=&\frac{2}{\sqrt{|g|}}\frac{\delta S}{\delta g^{\mu\nu}(x)} \label{eq:Tmn} \\
&=&\sum_{a=1}^N\partial_\mu \phi^a \partial_\nu\phi^a +\sum_{I,J=0}^{D-1}\partial_\mu X^I\partial_\nu X^J\eta_{IJ}-g_{\mu\nu} {\cal L},\label{eq:Tmn2}
\eea
and where the Lagrangian ${\cal L}$ is defined by the action in Eq.~(\ref{eq:S1}), $S\equiv\int d^Dx\,\sqrt{|g|}{\cal L}$.  An explicit solution to $T_{\mu\nu}(x)=0$ for the metric is
\begin{equation}
g_{\mu\nu}=\frac{D/2-1}{V(\phi^a)}\left(\sum_{a=1}^N\partial_\mu \phi^a\partial_\nu\phi^a+
\sum_{I,J=0}^{D-1}\partial_\mu X^I\partial_\nu X^J\eta_{IJ}\right),
\label{eq:gmn}\end{equation}
which allows the elimination of $g_{\mu\nu}$ from the action upon performing the ${\cal D} g_{\mu\nu}$ integration.  With the spacetime metric determined by Eq.~(\ref{eq:gmn}), the action for the theory resembles the  Dirac-Born-Infeld action with vanishing gauge field, modulated by the scalar-field potential function $V(\phi^a)$:
\begin{equation}
S=\int d^Dx\ \left(\frac{\tfrac D2-1}{V(\phi^a)} \right)^{\frac{D}{2}-1}
\sqrt {\bigg|\det \left(\sum_{a=1}^N \partial_\mu\phi^a \,\partial_\nu\phi^a 
+\sum_{I,J=0}^{D-1}\partial_\mu X^I \,\partial_\nu X^J\, \eta_{IJ}\right)\bigg|}.
\label{eq:S}\end{equation}
The general coordinate transformation invariance of Eq.~(\ref{eq:S1}) translated into reparametrization invariance of Eq.~(\ref{eq:S}). 
We proceed next to gauge fix by identifying the clock-and-ruler  fields $X^I$ with the corresponding spacetime coordinates, in analogy with the static gauge condition in string theory, up to an overall constant factor,
\begin{equation}
X^I=  x^\mu\delta_\mu^I \ \sqrt{\frac{V_0}{\tfrac{D}{2}-1}-c_1} , \ \ I=0,\dots,D-1 \, , \label{eq:staticgauge}
\end{equation}
where $c_1$ is a counterterm whose role will be revisited below.
In order to analyze the theory perturbatively,  we write $V(\phi)=V_0+\Delta V(\phi^a)$ and expand the action Eq.~(\ref{eq:S}) in powers of $1/V_0$. We also assume that $N$, the number of fields $\phi^a$ in the theory, is large and keep only leading terms in a $1/N$ expansion.  In the two-into-two scattering calculation of Ref.~\cite{Carone:2016tup}, this made the desired
diagrammatic resummation possible. 

The gauge-fixed action, expanded to second order in $1/V_0$, reads:
\begin{eqnarray}
S=\int d^Dx&&\left\{\frac{V_0}{D/2-1}+\frac{1}{2}\sum_{a=1}^N \partial_\mu \phi^a \partial^\mu \phi^a -\Delta V(\phi^a )\right. \nonumber \\ &&
-\frac{\tfrac D2-1}{4V_0}\left[
\sum_{a=1}^N\partial_\mu\phi^a \partial_\nu\phi^a 
\,\sum_{b=1}^N\partial^\mu\phi^b\partial^\nu \phi^b
-\frac{1}{2}\left(\sum_{a=1}^N \partial_\mu \phi^a \partial^\mu \phi^a  \right)^{\!\!2\,}\right] \nonumber \\
&& \left.-\frac{\tfrac{D}{2}-1}{2}\frac{\Delta V(\phi^a)}{V_0}\sum_{a=1}^N \partial_\mu \phi^a\partial^\mu \phi^a+\frac{D}{4}\frac{(\Delta V(\phi^a))^2}{V_0}+{\cal O}\left(\frac{1}{V_0^2}\right)\right\}.   \label{eq:Sexpansion}
\end{eqnarray}
We further assume a free theory with O$(N)$-symmetric potential 
\begin{equation}
\Delta V(\phi^a)=\sum_{a=1}^N\frac{m^2}{2}\phi^a\phi^a-c_2,\label{c_2}.  
\end{equation}

The counterterm $c_1$ from Eq.~(\ref{eq:staticgauge}) is chosen is to effectively normal order every occurence of $\partial_\mu \phi^a \partial_\nu\phi^a$ in Eq.~(\ref{eq:Sexpansion}) ({\em i.e.}, any loop which can be constructed by contracting the two $\phi^a$'s in $\partial_\mu \phi^a \partial_\nu\phi^a$ is rendered zero by adding the 
counterterm $-c_1 \eta_{\mu\nu} $). Also, the role of the counterterm $c_2$ from Eq.~(\ref{c_2}) is to normal order every occurrence of $\phi^a \phi^a$ ({\em i.e.}, any loop which can be constructed by contracting the two $\phi^a$'s in $m^2 \phi^a\phi^a$ is canceled by the counterterm $c_2$).

The graviton pole was identified in Ref.~\cite{Carone:2016tup} by considering the two-to-two scattering of $(\phi^a\, \phi^a\rightarrow \phi^b\,\phi^b) $ scalars in the large-$N$ limit\footnote{This approach is similar to that employed by Suzuki~\cite{suzuki} who identified a composite gauge boson in a particular scattering process in a theory of emergent electromagnetism.} as shown in Fig.~\ref{fig:match1}. Dimensional regularization was used as a regulator of the loop integrals. The existence of a massless spin-two state (the graviton) being exchanged in this process
required the fine-tuned choice\footnote{Here we correct a sign in the kernel $K^{\lambda\kappa}{}_{\rho\sigma}$ 
defined in Eq.~(3.16) of Ref.~\cite{Carone:2016tup}.  With $\lambda$ defined in the same way as in Ref.~\cite{Carone:2016tup}, this sign changes the
condition for the existence of a pole to $1+\lambda=1+ N(D/2-1)\Gamma(-D/2) (m^2)^{D/2}/(2 V_0 (4\pi)^{D/2})=0$, yielding Eq.~(\ref{choice}) above.}
\bea
V_0=-\frac{N(D/2-1)}{2} \frac{\Gamma(-D/2)}{(4\pi)^{D/2}} (m^2)^{D/2}    \,\, , \label{choice}
\eea
leading to the following expression for the scattering amplitude:
\begin{equation}
A^{\mu\nu|\rho\sigma}(q) =  -\frac{3 \,m^2}{D\, V_0}\, \left[(\tfrac D2-1) \,( \eta^{\nu\rho} \eta^{\mu\sigma} + \eta^{\nu\sigma} \eta^{\mu\rho}) 
- \eta^{\mu\nu} \eta^{\rho\sigma} \right] \, \frac{1}{q^2} +\cdots \,\,\, ,
\label{eq:ampsol}
\end{equation}
where we correct a sign error in Eq.~(3.20) of Ref.~\cite{Carone:2016tup}, which propagated into Eq.~(3.22) of that paper, and where the $\dots$ 
indicate terms which vanish as $q$ goes on-shell ({\em i.e.}, $q^2\to0$).  The definition of $A^{\mu\nu|\rho\sigma}(q)$ is reviewed in the next section.
Comparison with the corresponding graviton-mediated scattering amplitude in a free scalar theory\footnote{Here we correct yet another sign in the scattering amplitude given by Eq.~(3.26) of \cite{Carone:2016tup}.}
\begin{equation}
A^{\mu\nu|\rho\sigma}(q) = -\frac{M_{\Pl}^{2-D}}{D-2} \, \left[(\tfrac D2-1) \,( \eta^{\nu\rho} \eta^{\mu\sigma} + \eta^{\nu\sigma} \eta^{\mu\rho}) 
- \eta^{\mu\nu} \eta^{\rho\sigma} \right] \, \frac{1}{q^2} \,\,\, ,
\end{equation}
where $M_{\Pl}$ is the $D$-dimensional Planck mass, 
leads to
\begin{equation}
M_{\Pl}= m\,\bigg[\frac{N \, \Gamma(1-\frac{D}{2})}{6\, (4 \pi)^{D/2} }\bigg]^{1/(D-2)} \,\,\, . \label{eq:mpscat}
\end{equation}
With $D=4-\epsilon$, positivity of the Planck mass requires the regulator $\epsilon$ to be small and negative. The dimensionful constant $V_0$ as identified in Eq.~(\ref{choice}) is 
however positive.  For completeness we revisit the gauge choice Eq.~(\ref{eq:staticgauge}). Since $c_1$ is defined such that $c_1\eta_{\mu\nu}$ cancels any loop originating 
from  $\partial_\mu\phi^a \partial_\nu\phi^a$, expressing $c_1$ in terms of $V_0$ leads to 
\bea
c_1= \frac{V_0}{\tfrac D2-1},
\eea 
which differs by a sign relative to its expression given in footnote 6 of  Ref.~\cite{Carone:2016tup}.  Interestingly, the value of $c_1$ that eliminates 
tadpole diagrams exactly cancels the tree-level gauge choice for the $X^I$ in Eq.~(\ref{eq:staticgauge}).

\section{Contributions to the three-graviton coupling} \label{sec:contribs}

\begin{figure}[t]
  \begin{center}
    \includegraphics[width=.7\textwidth]{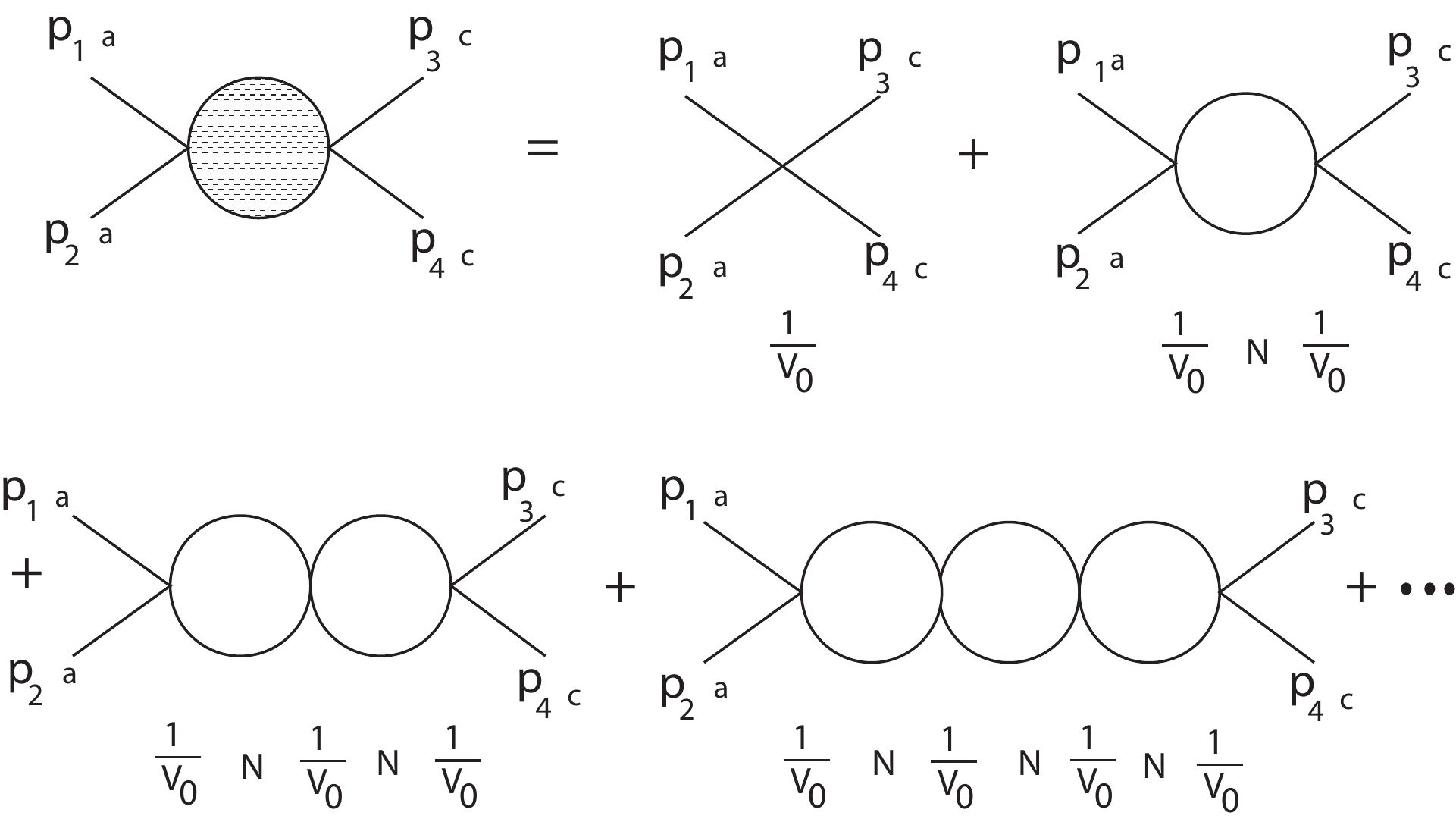}
    \caption{The scattering amplitude considered in Ref.~\cite{Carone:2016tup}}
        \label{fig:match1}
  \end{center}
\end{figure}

In Ref.~\cite{Carone:2016tup}, the existence of a propagating graviton was demonstrated non-perturbatively, at leading order in 
a $1/N$ expansion, by considering the infinite set of diagrams shown in Fig.~\ref{fig:match1}.  
The scattering amplitude was written
\begin{equation}
i {\cal M} (p_1,a \,; p_2,a \rightarrow p_3,c\,; p_4,c)\equiv E_{\mu\nu}(p_1,p_2) [i \,A^{\mu\nu|\rho\sigma}(q) ] \,E_{\rho\sigma}(p_3,p_4)\,\,\, ,
\end{equation}
with 
\begin{equation}
E_{\mu\nu}(p_1,p_2) \equiv - (p_1^\mu \, p_2^\nu + p_1^\nu \, p_2^\mu) + \eta^{\mu\nu} (p_1 \cdot p_2 +m^2) \,\,\, ,
\label{eq:exln}
\end{equation}
and where $q=p_1+p_2=p_3+p_4$.   The factor $E_{\mu\nu}$ corresponds to the Feynman rule for the flat-space
energy-momentum tensor for the noninteracting scalar fields, 
\begin{equation}
{\cal T}^{\mu\nu}= \sum_{a=1}^N\bigg[\partial^\mu \phi^a \partial^\nu \phi^a - \eta^{\mu\nu} \left(\frac{1}{2} \partial^\alpha \phi^a \partial_\alpha \phi^a - \frac{1}{2} m^2 \phi^a \phi^a \right)\bigg]  \,\,\, .
\label{eq:tflat}
\end{equation}
The gauge-invariant part of the graviton propagator was extracted by studying the $1/q^2$ pole in 
$A^{\mu\nu|\rho\sigma}(q)$, and was shown to have the appropriate form.   The calculation depended only on the quartic 
scalar interaction vertex in the theory, 
\begin{equation}
{\cal L}_{int} = -\frac{1}{4 V_0} {\cal T}_{\mu\nu} \, {\cal T}_{\alpha\beta} \, P^{\mu\nu|\alpha\beta} \,\,\, ,
\label{eq:int1}
\end{equation}
where we defined the tensor structure
\begin{equation}
 P^{\alpha\beta|\lambda\kappa} \equiv \frac{1}{2}  \left[ (\tfrac D2-1)\left(\eta^{\alpha\lambda} \eta^{\beta\kappa} + \eta^{\alpha\kappa} \eta^{\beta\lambda}\right)
 - \eta^{\lambda\kappa}\eta^{\alpha\beta} \right] \,\,\, . \label{eq:P}
 \end{equation}
At either end of the sum of scattering diagrams in Fig.~\ref{fig:match1}, there are two possible ways of choosing the 
${\cal T}_{\kappa\lambda}$ factor in Eq.~(\ref{eq:int1}) that corresponds to the external scalar lines;  if we include this combinatoric factor, 
we may write an effective interaction that represents the coupling of the graviton to two scalars,
\begin{equation}
{\cal L}_{eff} = -\frac{1}{2} \, h^{(1)}_{\alpha\beta} \, {\cal T}^{\alpha\beta}   \,\,\, ,
\label{eq:thv}
\end{equation}
\begin{figure}[t]
  \begin{center}
    \includegraphics[width=.4\textwidth]{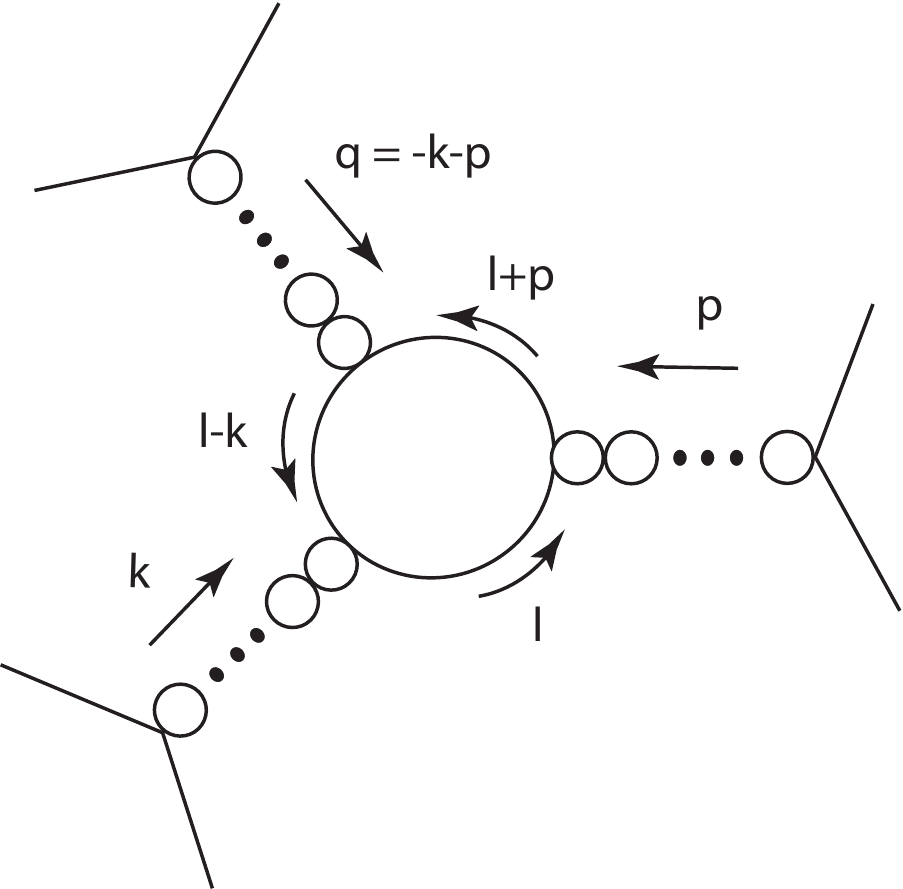}
    \caption{Contribution to the loop-generated three-graviton coupling involving the quartic scalar interaction vertex exclusively.  The 
    chains of small circles corresponds to the sum over loops defined in Fig.~\ref{fig:match1}, which leads to graviton poles.  Each 
    pair of external scalars is assumed to be distinct.}    
        \label{fig:scatdiag1}
  \end{center}
\end{figure}
where $h^{(1)}_{\alpha\beta} = {\cal T}^{\mu\nu} P_{\mu\nu|\alpha\beta} / V_0$.  (The meaning 
of the superscript will be explained below.)   Eq.~(\ref{eq:thv}) is of the same form as the graviton-scalar-scalar 
coupling that one would find in a linearized theory with a fundamental graviton field, up to the normalization of the field.  However, 
$-h^{(1)}_{\alpha\beta}/2$  here represents a composite operator that is quadratic in the scalar fields, with a two-point
correlation function given by $i A^{\mu\nu|\rho\sigma}(q)$.  In Ref.~\cite{Carone:2016tup}, this amplitude was found
to contain a pole in $q^2$ proportional to $i \, P_{\alpha\beta | \mu\nu} / M_\Pl^2$ in $D=4$.  Thus, one contribution 
to the emergent three-graviton coupling arises from a central scalar loop that connects three factors of $T^{\mu\nu}$ 
associated with the effective interaction in Eq.~(\ref{eq:thv}).   This is illustrated in the six-scalar scattering amplitude shown
in Fig.~\ref{fig:scatdiag1}.  The sum of loops representing each ``leg" of this diagram was evaluated in Ref.~\cite{Carone:2016tup} 
at leading order in an expansion in $q^2$ and $1/N$.  Using this result, we can isolate the leading part of the amplitude 
in Fig.~\ref{fig:scatdiag1} in a similar expansion.   After including the additional contributions described below, the final 
scattering amplitude can be compared to the same quantity computed in general relativity.

\begin{figure}[t]
  \begin{center}
    \includegraphics[width=.7\textwidth]{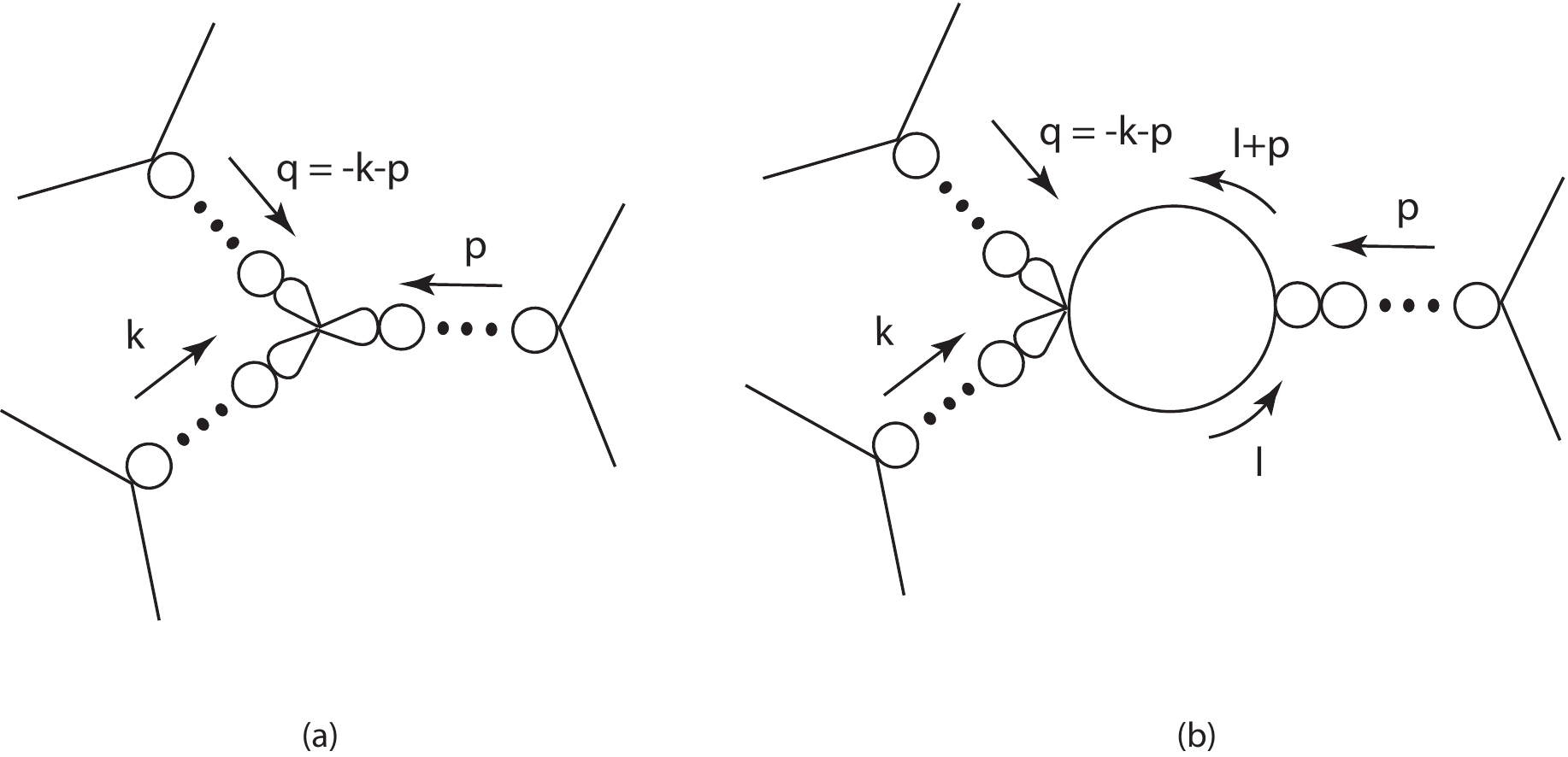}
    \caption{Contribution to the loop-generated three-graviton coupling involving six-scalar interaction vertices. The 
    chains of small circles corresponds to the sum over loops defined in Fig.~\ref{fig:match1}, which leads to graviton poles.  Each 
    pair of external scalars is assumed to be distinct.  Two other diagrams similar to the second one are not shown.}    
        \label{fig:scatdiag2}
  \end{center}
\end{figure}

Additional contributions to the emergent three-graviton coupling arise from the presence of six-scalar interactions 
at the next order in $1/V_0$.  Using the solution for $g_{\mu\nu}$ that yields a vanishing total energy-momentum tensor, we 
may write $g_{\mu\nu} = \eta_{\mu\nu} + h_{\mu\nu}$, where
\begin{equation}
h_{\mu\nu} = h^{(1)}_{\mu\nu} + h^{(2)}_{\mu\nu}+ \cdots  \,\,\, ,
\end{equation}
with
\begin{equation}
h^{(1)}_{\mu\nu} = -\frac{\Delta V}{V_0} \, \eta_{\mu\nu} + \frac{(D/2-1)}{V_0} \partial_\mu \phi \cdot \partial_\nu \phi 
\, \equiv \, \frac{1}{V_0}\, {P^{\alpha\beta}}_{\lambda\kappa}\, {\cal T}_{\alpha\beta}  \,\,\, ,
\label{eq:h1exp}
\end{equation}
and
\begin{equation}
h^{(2)}_{\mu\nu} = - \frac{\Delta V}{V_0} \, h^{(1)}_{\mu\nu} \,\,\, .
\label{eq:h2exp}
\end{equation}
The second equality in Eq.~(\ref{eq:h1exp}) follows after some simple algebra. The operator $h^{(1)}_{\mu\nu}$ is the 
same as the one identified in our discussion of the graviton-scalar-scalar coupling.  The superscript indicates the order in an expansion in 
$1/V_0$. Using Eqs.~(\ref{eq:h1exp}) and (\ref{eq:h2exp}), we expand the action in our theory, Eq.~(\ref{eq:S}), to order $1/V_0^2$, while expressing the result entirely 
in terms of $h^{(1)}_{\mu\nu}$:
\begin{align}
{\cal L}_{int}  = \frac{V_0}{(D/2-1)} & \left[\frac{1}{6} \, {{h^{(1)}}^\alpha}_\beta \, {{h^{(1)}}^\beta}_\gamma \, {{h^{(1)}}^\gamma}_\alpha - \frac{1}{8} \, h^{(1)} 
{{h^{(1)}}^\alpha}_\beta {{h^{(1)}}^\beta}_\alpha  + \frac{1}{48} \, {h^{(1)}}^3 \right. \nonumber \\
& \left. -\frac{\Delta V}{V_0} \left(-\frac{1}{4} {{h^{(1)}}^\alpha}_\beta {{h^{(1)}}^\beta}_\alpha + \frac{1}{8} {h^{(1)}}^2 \right) \right] \,\,\, .
\label{eq:2gstart}
\end{align}
Each factor of $-h^{(1)}/2$ in Eq.~(\ref{eq:2gstart}) can connect to the sum over loop diagrams, described earlier, that generate graviton poles; the first three terms in Eq.~(\ref{eq:2gstart}) lead to an effective three-graviton contact interaction that is independent of momentum.   No such interaction exists in general relativity; we will show later how the effects of these terms are cancelled.   A separate contribution to the three-graviton coupling arises in a similar way from the remaining terms in Eq.~(\ref{eq:2gstart}), except that the factor of 
$\Delta V$ must be joined to the quartic scalar vertex in Eq.~(\ref{eq:thv}) by a loop.  The contributions to the six-scalar scattering amplitude discussed earlier that follow 
from the interactions in Eq.~(\ref{eq:2gstart}) are shown in Fig.~\ref{fig:scatdiag2}.
Note that a contribution to the three-graviton coupling arising from the eight-scalar interactions, at order $1/V_0^3$, 
is absent in this theory.  Such a diagram would require closing a single scalar loop that is cancelled by the counterterms, appearing in Eqs.~(\ref{eq:staticgauge}) and
(\ref{c_2}), that remove all the tadpole diagrams in the theory.

We evaluate the diagrams of Figs.~\ref{fig:scatdiag1} and \ref{fig:scatdiag2} in the next section.  We show that the part of the result that is 
quadratic in the graviton momenta is consistent with the three-graviton coupling of general relativity, and the part that is independent of momentum is vanishing.
We verify that the Planck scale determined from this calculation agrees with the result obtained via the scattering amplitude of Ref.~\cite{Carone:2016tup}.   In both 
calculations, loops are regulated using dimensional regularization, with $D=4-\epsilon$, and $\epsilon$ fixed at a {\em finite} value.   This approach is 
used as a placeholder for whatever generally covariant physical regulator may be provided by a realistic ultraviolet completion of the theory.   Proceeding in this 
way, the Planck scale becomes a function of $1/\epsilon$ and we will verify that the result obtained in the next section agrees 
exactly with the one found in Ref.~\cite{Carone:2016tup}.   

\section{Emergent three-graviton vertex and $M_\Pl$} \label{sec:3gmp}

In this section, we study the six-scalar scattering amplitude given by the diagrams shown in Figs.~\ref{fig:scatdiag1} and \ref{fig:scatdiag2}, and compare to the result
expected from general relativity.   The chains of circles forming the three legs of the diagrams represent the same sums over loops shown in Fig.~\ref{fig:match1} that 
provide the emergent graviton poles.   Since these were previously isolated as the leading order terms in an expansion in $q^2$, we can reliably compute the leading 
order piece of the scattering amplitude as the graviton legs are taken nearly on shell.  It is convenient to parameterize the result as
\begin{equation}
i {\cal M} =  \frac{1}{M_\Pl^6} \frac{1}{p^2}\frac{1}{q^2} \frac{1}{k^2} \, {\cal A}  \,\,\, ,
\end{equation}
where we define
\begin{equation}
{\cal A} = E_1^{\lambda\kappa} \, E_2^{\psi\omega} \, E_3^{\phi\chi} \, P_{\lambda\kappa | \alpha\beta} \,
P_{\psi\omega | \gamma\delta} \, P_{\phi\chi | \epsilon\zeta}  \, A^{\alpha\beta\gamma\delta\epsilon\zeta}  \,\,\, .
\label{eq:adef}
\end{equation}
Here, $E_1$, $E_2$ and $E_3$ refer to the external line factors, defined earlier, that connect to graviton lines carrying momenta $p$, $k$ and $q$, respectively.
Note that we will only need to evaluate the $1/\epsilon$ dependence of the central loops in the diagrams of Fig.~\ref{fig:scatdiag1} and \ref{fig:scatdiag2}, so that all 
vertices are evaluated for $D=4$.   

The inner-most loop in the diagram of Fig.~\ref{fig:scatdiag1} involves three insertions of the energy-momentum tensor ${\cal T}^{\mu\nu}$, which has the Feynman rule 
\begin{equation}
i \, \left[ p_1^\mu \, p_2^\nu + p_1^\nu \, p_2^\mu + \eta^{\mu\nu} (m^2 - p_1 \cdot p_2) \right]  \,\,\, ,
\end{equation}
for momentum $p_1$ flowing in and momentum $p_2$ flowing out.  The amplitude is then given by
\begin{equation}
A^{\alpha\beta\gamma\delta\epsilon\zeta}_1 = N \, \int \frac{d^4 \ell}{(2 \pi)^4}   \frac{N^{\alpha\beta\gamma\delta\epsilon\zeta}_1}{
[(p+\ell)^2-m^2] [\ell^2-m^2][(\ell-k)^2-m^2]} \,\,\, ,
\label{eq:amp1}
\end{equation}
where the numerator factor is
\begin{align}
N^{\alpha\beta\gamma\delta\epsilon\zeta}_1 & = [\ell^\alpha(\ell+p)^\beta+\ell^\beta(\ell+p)^\alpha+\eta^{\alpha\beta} (m^2 - \ell \cdot (\ell+p))] \nonumber \\
& \times [(\ell+p)^\epsilon(\ell-k)^\zeta+(\ell+p)^\zeta(\ell-k)^\epsilon+\eta^{\epsilon\zeta}(m^2-(\ell+p)\cdot(\ell-k))] \nonumber \\
&\times [(\ell-k)^\gamma \ell^\delta + (\ell-k)^\delta \ell^\gamma+\eta^{\gamma\delta}(m^2-(\ell-k)\cdot \ell)]  \,\,\, ,
\end{align}
and momenta are defined as in Fig.~\ref{fig:scatdiag1}.  Similarly, diagram of Fig.~\ref{fig:scatdiag2}b leads to the amplitude
\begin{equation}
A^{\alpha\beta\gamma\delta\epsilon\zeta}_2 = m^2 \,  N \, \int \frac{d^4 \ell}{(2 \pi)^4}  \frac{N^{\alpha\beta\gamma\delta\epsilon\zeta}_2}{
[(p+\ell)^2-m^2] [\ell^2-m^2]} \,\,\, ,
\label{eq:amp2}
\end{equation}
where the numerator factor is 
\begin{equation}
N^{\alpha\beta\gamma\delta\epsilon\zeta}_2 = 
- \left[ \ell^\alpha (\ell+p)^\beta + \ell^\beta (\ell+p)^\alpha + \eta^{\alpha\beta} (m^2 - \ell \cdot (\ell+p)) \right] P^{\gamma \delta | \epsilon\zeta}
\,\,\,. 
\end{equation}
Let us first focus on the part of the amplitude that is quadratic in the momenta $p$, $k$ and $q$, which excludes any contribution from the diagram of
Fig.~\ref{fig:scatdiag2}a.   We note by inspection of Eq.~(\ref{eq:amp2}) that $A^{\alpha\beta\gamma\delta\epsilon\zeta}_2$ will involve terms
proportional to either $p^\alpha p^\beta$ or $p^2 \, \eta^{\alpha\beta}$.  Terms of the first type vanish since they are contracted with the external line
factor $E_1$, while terms of the second type are higher-order in our expansion about the limit $p^2=k^2=q^2=0$.   Hence, the diagram of Fig.~\ref{fig:scatdiag2}b
does not contribute at leading order to the part of the amplitude ${\cal A}$ that is quadratic in the graviton momenta.

The diagram of Fig.~\ref{fig:scatdiag1}, however, does contribute.  Using Eq.~(\ref{eq:adef}) and 
the notation ${{E_i}^\mu}_\mu \equiv E_i$ and $ {E_i}_{\mu\nu} E_j^{\mu\nu} \equiv E_i \cdot E_j$, we find
\begin{align}
{\cal A}  =& \frac{i N}{8 \pi^2} \frac{m^2}{\epsilon} \left[-\frac{1}{6} \, E_1 E_2 E_3^{\mu\nu}k_\mu k_\nu
+\frac{1}{3}\, E_1 \cdot E_2 \, E_3^{\mu\nu}k_\mu k_\nu \right.\nonumber \\
&\left.
 -\frac{2}{3} \,E_1^{\mu\nu} k_\nu E_2^{\alpha\beta} p_\beta \, {E_3}_{\mu\alpha}  
+ \mbox{ perms } \right] + \mbox{ higher order }  \,\,\, ,
\label{eq:a1}
\end{align}
where ``perms" refers to two additional cyclic permutations in which 
\begin{equation}
(E_1, \, p) \rightarrow (E_2,\, k) \rightarrow (E_3, \, q) \rightarrow (E_1, \, p)\,\,\, ,
\end{equation}
and ``higher order" represents terms that vanish when the gravitons are exactly on shell.

\begin{figure}[t]
  \begin{center}
    \includegraphics[width=.4\textwidth]{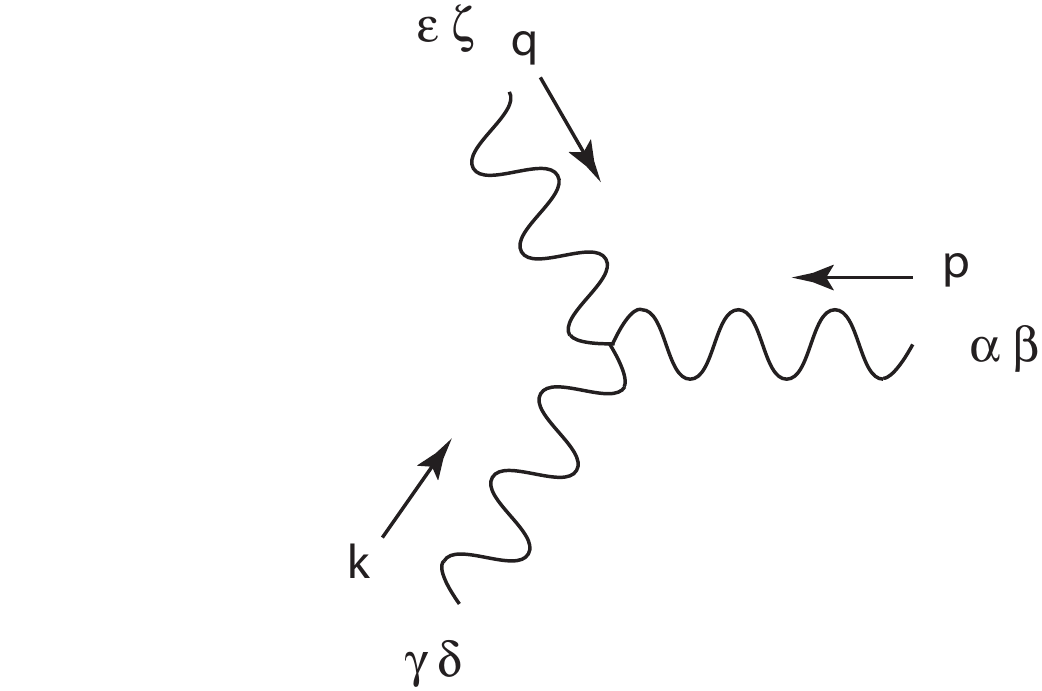}
    \caption{Conventions for the Feynman rule for the three-graviton vertex.}
        \label{fig:3gv}
  \end{center}
\end{figure}

We can compute the analogous result in the weak field expansion of general relativity.  We work with a rescaling
of the graviton field so that the propagator in de Donder gauge is $i \, P_{\alpha\beta | \mu\nu} / q^2$, and the coupling of the graviton to the 
energy-momentum tensor is $h_{\mu\nu} {\cal T}^{\mu\nu}/M_\Pl$.   Defining the following quantities,
\begin{align}
f_1  &= -  \, i\,  p^2 \, \eta_{\alpha\beta}\, \eta_{\gamma \delta}  \, \eta_{\epsilon\zeta} \,\,\, ,  \\
f_2  &= - \frac{i}{2} \, q^2 \, \eta_{\alpha \beta} (\eta_{\gamma\zeta} \, \eta_{\epsilon\delta} + \eta_{\delta\zeta} \, \eta_{\epsilon\gamma}) \,\,\, ,\\
f_3  &= -\frac{i}{2}\, q \cdot k  \, \eta_{\alpha \beta} (\eta_{\gamma\epsilon}\,\eta_{\delta\zeta} + \eta_{\epsilon\delta} \, \eta_{\zeta\gamma}) \,\,\, , \\
f_4  &=- \frac{i}{8}\, q^2 \, ( \eta_{\beta\epsilon}\,\eta_{\gamma\zeta}\, \eta_{\alpha\delta} +\eta_{\beta\epsilon}\,\eta_{\delta\zeta}\, \eta_{\alpha\gamma}
+\eta_{\beta\zeta}\,\eta_{\gamma\epsilon}\, \eta_{\alpha\delta} +\eta_{\beta\zeta}\,\eta_{\delta\epsilon}\, \eta_{\alpha\gamma} \nonumber \\
&+\eta_{\alpha\epsilon}\,\eta_{\gamma\zeta}\, \eta_{\beta\delta} +\eta_{\alpha\epsilon}\,\eta_{\delta\zeta}\, \eta_{\beta\gamma} 
+\eta_{\alpha\zeta}\,\eta_{\gamma\epsilon}\, \eta_{\beta\delta} +\eta_{\alpha\zeta}\,\eta_{\delta\epsilon}\, \eta_{\beta\gamma} ) \,\,\, ,\\
f_6 &= - \, i \, k_\alpha k_\beta \, \eta_{\epsilon\zeta} \, \eta_{\gamma\delta}  \,\,\, , \\
f_7 &= -\frac{i}{4} \, (q_\alpha k_\beta + q_\beta k_\alpha) \, (\eta_{\epsilon\delta} \, \eta_{\zeta\gamma} + \eta_{\epsilon\gamma} \, \epsilon_{\zeta\delta} )  \,\,\, \\
f_8 &= -\frac{i}{2}  \, k_\alpha k_\beta (\eta_{\epsilon\gamma} \, \eta_{\zeta\delta} + \eta_{\epsilon \delta} \, \eta_{\zeta \gamma}) \,\,\, , \\
f_9 &= -\frac{i}{4}\, \eta_{\alpha\beta} (q_\epsilon k_\gamma \eta_{\delta\zeta} + q_\epsilon k_\delta \eta_{\gamma \zeta} + q_\zeta k_\gamma \eta_{\delta\epsilon}
+q_\zeta k_\delta \eta_{\gamma\epsilon}) \,\,\, ,\\
f_{10} &= -\frac{i}{4}\, \eta_{\alpha\beta} (q_\epsilon q_\gamma \eta_{\zeta\delta}+q_\epsilon q_\delta \eta_{\zeta\gamma}+q_\zeta q_\gamma \eta_{\epsilon\delta}
+q_\zeta q_\delta \eta_{\epsilon\gamma}) \,\,\, , \\
f_{11} &= -\frac{i}{4} \, \eta_{\alpha\beta} (k_\epsilon q_\gamma \eta_{\zeta\delta} + k_\epsilon q_\delta \eta_{\zeta \gamma}
+k_\zeta q_\gamma \eta_{\epsilon\delta} +k_\zeta q_\delta \eta_{\epsilon\gamma}) \,\,\,, \\
f_{13} &= -\frac{i}{8} \, (q_\epsilon q_\gamma \eta_{\alpha\zeta} \, \eta_{\beta\delta} +q_\epsilon q_\delta \eta_{\alpha\zeta} \, \eta_{\beta\gamma}
+q_\zeta q_\gamma \eta_{\alpha\epsilon} \, \eta_{\beta\delta} +q_\zeta q_\delta \eta_{\alpha\epsilon} \, \eta_{\beta\gamma} \nonumber \\
&+q_\epsilon q_\gamma \eta_{\beta\zeta} \, \eta_{\alpha\delta} +q_\epsilon q_\delta \eta_{\beta\zeta} \, \eta_{\alpha\gamma}
+q_\zeta q_\gamma \eta_{\beta\epsilon} \, \eta_{\alpha\delta} +q_\zeta q_\delta \eta_{\beta\epsilon} \, \eta_{\alpha\gamma}) \,\,\, , \\
f_{14} &= - \, \frac{i}{8} \, [ (k_\epsilon \eta_{\alpha\zeta} + k _\zeta \eta_{\alpha\epsilon})(q_\gamma \eta_{\beta\delta} + q_\delta \eta_{\beta\gamma})
+(k_\epsilon \eta_{\beta\zeta} + k _\zeta \eta_{\beta\epsilon})(q_\gamma \eta_{\alpha\delta} + q_\delta \eta_{\alpha\gamma})]  \,\,\,,
\end{align}
we find that the Feynman rule for the three-graviton vertex shown in Fig.~\ref{fig:3gv} can be written
\begin{equation}
i \, V_{3h}^{\alpha\beta\gamma\delta\epsilon\zeta} = - \frac{4}{M_\Pl} \, \left(\sum_i c_i f_i + \mbox { perms } \right) \,\,\, ,
\label{eq:3hmess}
\end{equation}
where ``perms"  refers to the five remaining permutations of the labels of the external lines $(\alpha,\beta,p)$, $(\gamma,\delta,k)$, and 
$(\epsilon,\zeta,q)$.  The coefficients $c_i$ are given by
\begin{equation}
\begin{array}{lllllll}
c_1=\frac{1}{16}, & c_2=-\frac{1}{2},  & c_3=-\frac{3}{8},  & c_4=\frac{1}{4},  & c_5=0 , & c_6= -\frac{1}{4},
& c_7 = \frac{1}{4},   \\
c_8 = \frac{1}{2}, & c_9=\frac{1}{2} , & c_{10}=1, & c_{11}=\frac{1}{4},  & c_{12}=0,
& c_{13}=-1 , & c_{14}=-\frac{1}{2} \, .
\end{array} 
\end{equation}
The origin of the $f_i$ and the normalization of Eq.~(\ref{eq:3hmess})  are described briefly in Appendix~\ref{sec:appendix}\footnote{We would like to warn the attentive reader that the three-graviton vertex as given in Feynman's lectures on gravitation \cite{feynman} is incomplete. In the literature there are more succint expressions for the vertex, but they refer to the interactions of another field, namely the tensor density
${\mathfrak g}^{\mu\nu}\equiv \sqrt{-g} g^{\mu\nu} = \eta^{\mu\nu}+{\mathfrak h}^{\mu\nu}$. See for example
\cite{goldberg} and  \cite{duff}.} Computing the same six-scalar 
amplitude that we considered in our purely scalar theory, we find 
\begin{align}
{\cal A}_{GR}  &= - 6\, i \, M_\Pl^2 \left[(k^2 + p^2 + q^2)
(\frac{1}{3} \, E_1^{\mu\nu} \, {{E_2}_\nu}^\alpha \, {E_3}_{\alpha\mu} +\frac{1}{12} \, E_1 E_2 E_3) \right. \nonumber \\
& + \left\{-\frac{1}{6} \, E_1 E_2 E_3^{\mu\nu}k_\mu k_\nu
+\frac{1}{3}\, E_1 \cdot E_2 \, E_3^{\mu\nu}k_\mu k_\nu -\frac{2}{3} \,E_1^{\mu\nu} k_\nu E_2^{\alpha\beta} p_\beta \, {E_3}_{\mu\alpha}  \right. \nonumber \\
& \left. \left. -\frac{1}{6}\,(p^2+k^2) E_3 \, E_1\cdot E_2 + \mbox{ perms }\right\} \right] \,\,\, .
\label{eq:grres}
\end{align}
In the limit where the gravitons are nearly on shell, this result coincides with the amplitude ${\cal A}$ determined in our scalar theory provided we identify 
$- 6\, i \, M_\Pl^2$ with $i N m^2 / ( 8 \pi^2\epsilon)$.  Hence, we conclude 
\begin{equation}
M_\Pl^2 = - \frac{N}{48 \pi^2} \frac{m^2}{\epsilon} \,\,\,,
\end{equation}
with $\epsilon<0$.   Although the $1/\epsilon$ originates here from a loop diagram in the three-graviton vertex calculation,
the resulting relation to $M_\Pl$ agrees with the scattering calculation of Ref.~\cite{Carone:2016tup}, {\em i.e.}, Eq.~(\ref{eq:mpscat}).

We now consider the momentum-independent contributions to the amplitude ${\cal A}$, which we denote ${\cal A}^0$.    The contact terms shown in Fig.~\ref{fig:scatdiag2}a give
the contribution
\begin{equation}
{\cal A}^0_{tree} = - i V_0 \left[ - E_1 E_2 E_3 + 2 (E_1 E_2 \cdot E_3 + \mbox{ perms })- 8\, E_1^{\mu\nu} \, {{E_2}_\nu}^\alpha \, {E_3}_{\alpha\mu} \right] \,\,\, . \label{eq:bad1}
\end{equation}
The central loop in Fig.~\ref{fig:scatdiag2}b gives no contribution; this is clear since a momentum-independent part would have to remain in the limit $p^\mu=0$, and we
have already established that this contribution to ${\cal A}$ exclusively involves terms proportional to $p^\alpha \, p^\beta$ or $p^2 \, \eta^{\alpha\beta}$.  On the other hand, we may evaluate
the result coming from the diagram of Fig.~\ref{fig:scatdiag1}.  We find
\begin{equation}
{\cal A}^0_{loop} = -\frac{1}{2} \frac{i N}{16 \pi^2} \frac{m^4}{\epsilon} \left[ - E_1 E_2 E_3 + 2 (E_1 E_2 \cdot E_3 + \mbox{ perms })- 8\, E_1^{\mu\nu} \, {{E_2}_\nu}^\alpha \, {E_3}_{\alpha\mu} \right] \,\,\, . \label{eq:bad2}
\end{equation}
Cancellation is assured if 
\begin{equation}
V_0 = -\frac{N}{2} \frac{m^4}{16 \pi^2 \epsilon}  \,\,\, ,
\end{equation}
which is satisfied for the same value of $V_0$ that gave us a massless graviton pole.  Aside from higher-derivative corrections, our results for the three-graviton coupling 
are consistent with the expectations of general relativity.

This conclusion is gratifying since the action of our theory is generally covariant and the choice of the background for the clock-and-ruler fields is a 
convenient gauge choice, analogous to static gauge in string theory.  We therefore expect that the form of the three-point coupling should be in accord 
with general relativity, aside from higher-derivative corrections.  Among the sign corrections discussed in Sec.~\ref{sec:model}, we note that the 
corrected sign in the expression for $V_0$ relative to the result in Ref.~\cite{Carone:2016tup} not only assures the masslessness of the graviton, 
but here also provides for the nontrivial cancellation between the contact terms in Eq.~(\ref{eq:bad1}), and the one-loop terms in Eq.~(\ref{eq:bad2}).

Finally, we comment on higher-derivative corrections.  The diagram in Fig.~\ref{fig:scatdiag1} also contributes to terms in the three-graviton vertex that 
involve four graviton momenta.  Since this part of the central loop is logarithmically divergent, the result is proportional to $N/\epsilon$, which
is suppressed by a factor of $m^2$ relative to the result of Eq.~(\ref{eq:a1}).   For large $N/\epsilon$, $m$ can be substantially smaller than $M_\Pl$,
but still sufficient to render these effects harmless at the distance scales where gravity has been probed~\cite{rppgrav}.  For the sake of argument, 
if $\epsilon \sim 10^{-11}$, consistent with the phenomenological bound $\epsilon < 4 \times 10^{-11}$ from Ref.~\cite{Schafer:1986oda}, and $N=100$, 
one finds $m \approx 7 \times 10^{-6} M_\Pl \approx 10^{13}$~GeV.   However, it should be stressed that the suppression of these higher-derivative 
interactions by $m^2$ rather than $M_\Pl^2$ is peculiar to dimensional regularization, where the regulator is dimensionless; this result does not
generalize to other regulators, for example, Pauli-Villars fields or a Schwinger time / heat kernel UV regulator.   We therefore do not consider the suppression of higher-derivative interactions by powers of the scalar mass to be an intrinsic feature of these models.

\section{Universality of the Graviton Coupling} \label{sec:uni}

In this section, we consider a theory with two distinct sets of noninteracting scalar fields, $\phi^a_1$ with $a=1 \ldots N_1$, 
and $\phi^b_2$ with $b=1 \dots N_2$, distinguished only by their masses $m_1$ and $m_2$.   We wish to show that extending the 
scalar theory in this way preserves a massless spin-2 graviton state, and that the graviton couples universally to both 
types of scalars.

The action of the theory, prior to imposing the constraint of vanishing total energy-momentum tensor, is
\begin{equation}
S = \int d^D x \sqrt{|g|} \left[\frac{1}{2} g^{\mu\nu} \left( \sum_{a=1}^{N_1} \partial_\mu \phi_1^a \partial_\nu \phi_1^a + \sum_{a=1}^{N_2} \partial_\mu \phi_2^a \partial_\nu \phi_2^a \right)
- \sum_{I,J=0}^{D-1} \partial_\mu X^I \partial_\nu X^J \eta_{IJ} - V(\phi_1,\phi_2) \right]  \,\,\, .
\end{equation}
This extension of the original theory is obtained via the replacements
\begin{equation}
 \sum_{a=1}^{N} \partial_\mu \phi^a \partial_\nu \phi^a \longrightarrow \sum_{a=1}^{N_1} \partial_\mu \phi_1^a \partial_\nu \phi_1^a + \sum_{a=1}^{N_2} \partial_\mu \phi_2^a \partial_\nu \phi_2^a  \,\,\, ,
\end{equation} 
and
\begin{equation}
V(\phi) \longrightarrow V(\phi_1,\phi_2) \equiv V_0 + \Delta V(\phi_1,\phi_2) \,\,\, ,
\end{equation}
where
\begin{equation}
\Delta V(\phi_1,\phi_2) = \frac{m_1^2}{2} \sum_{a=1}^{N_1}  \phi_1^a \, \phi_1^a + \frac{m_2^2}{2} \sum_{a=1}^{N_2}  \phi_2^a \, \phi_2^a  \,\,\, .
\end{equation}
It follows that
\begin{equation}
{\cal T}^{\mu\nu} = {\cal T}_1^{\mu\nu}+{\cal T}_2^{\mu\nu}  \,\,\, .
\end{equation}
The interaction vertex that is relevant to the two-into-two scattering calculation can be inferred from the results of Sec.~\ref{sec:contribs}
\begin{equation}
{\cal L}_{int} = -\frac{1}{4 \, V_0} P^{\alpha\beta | \mu\nu} ({\cal T}_1+{\cal T}_2)_{\mu\nu} 
\, ({\cal T}_1+{\cal T}_2)_{\alpha \beta} \,\,\,.
\label{eq:newLint}
\end{equation}
Generalizing our earlier notation, we define the functions
\begin{equation}
E_i^{\mu\nu}(p,k) \equiv -(p^\mu\, k^\nu + p^\nu\, k^\mu) + \eta^{\mu\nu} (\, p \cdot k + m_i^2\, )  \,\,\,,
\end{equation}
for $i=1,2$, corresponding to the Feynman rule for a ${\cal T}_i$ factor, assuming inwardly or outwardly directed momenta $p$ and $k$.   
\begin{figure}[t]
  \begin{center}
    \includegraphics[width=.6\textwidth]{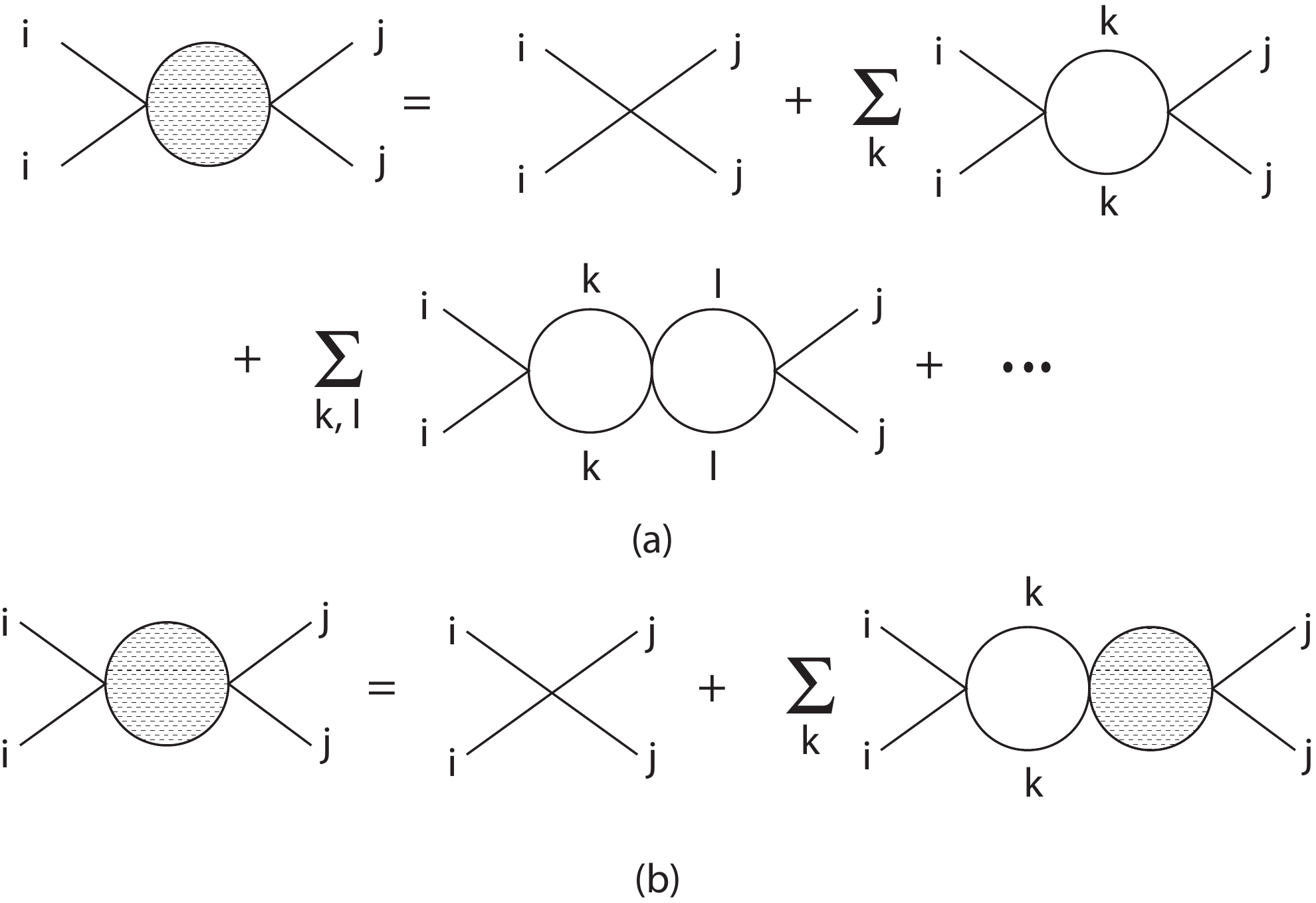}
    \caption{Two-into-two scattering diagrams.  The labels take the values $1$ or $2$, indicating the set of scalars to 
    which a given line is associated;  external lines are chosen such that only s-channel diagrams are relevant.}    
        \label{fig:uni}
  \end{center}
\end{figure}
The scattering amplitude shown in Fig.~\ref{fig:uni}a has the form
\begin{equation}
i {\cal M}(p_1, \, i \, ; \, p_2 , \, i \longrightarrow p_3,\,j \,;\, p_4,\, j) = E^{\mu\nu}_i (p_1,p_2) \left[ i\, A_{ij}(q)_{\mu\nu | \rho\sigma} \right]
E^{\rho\sigma}_j(p_3,p_4) \,\,\,.
\end{equation}
with $q=p_1+p_2=p_3+p_4$.  This is equivalent to the recursive representation shown in Fig.~\ref{fig:uni}b, which leads to
\begin{equation}
A^{\mu\nu|\rho\sigma}_{ij}={A_0}^{\mu\nu|\rho\sigma}_{ij} + \sum_k ({K_{i\,k})^{\mu\nu}}_{\alpha\beta} \, A^{\alpha\beta|\rho\sigma}_{kj}  \,\,\, ,
\label{eq:kernel1}
\end{equation}
where $A_0$ corresponds to the tree-level amplitude and we have stripped off the external line factors.  It follows from the form of the 
interaction in Eq.~(\ref{eq:newLint}), however, that the amplitudes $A^{\mu\nu|\rho\sigma}_{i,j}$ are independent of $i$ and $j$.  Dropping 
these labels, we may write
Eq.~(\ref{eq:kernel1}) as
\begin{equation}
A^{\mu\nu|\rho\sigma} = A_0^{\mu\nu|\rho\sigma} +  {K^{\mu\nu}}_{\alpha\beta} \, A^{\alpha\beta|\rho\sigma} \,\,\, ,
\label{eq:kernel2}
\end{equation}
where
\begin{equation}
{K^{\mu\nu}}_{\alpha\beta} = {K^{\mu\nu}}_{\alpha\beta}(N_1,m_1)+{K^{\mu\nu}}_{\alpha\beta}(N_2,m_2) \,\,\, ,
\end{equation}
with the kernel ${K^{\mu\nu}}_{\alpha\beta}(N, m)$ identical to that of a theory with a single set of $N$ scalar fields with masses $m$~\cite{Carone:2016tup},
\begin{equation}
{K^{\mu\nu}}_{\alpha\beta} = - \frac{N(D/2-1)}{4 V_0} \frac{\Gamma(-D/2)}{(4\pi)^{D/2}} (m^2)^{D/2} \left[1-\frac{D}{12} \frac{q^2}{m^2}\right]\left(\delta^\nu_\alpha \, 
\delta^\mu_\beta + \delta^\nu_\beta \,
\delta^\mu_\alpha\right) + {\cal O}(q^4)\,\,\, ,
\label{eq:kernalfixed}
\end{equation}
where we have included the sign correction noted in Sec.~\ref{sec:model}. The condition that the amplitude in Eq.~(\ref{eq:kernel2}) includes a massless pole generalizes to
\begin{equation}
V_0 = - \frac{(D/2-1)}{2} \frac{\Gamma(-D/2)}{(4 \pi)^{D/2}} \, \sum_i \, N_i \, (m_i^2)^{D/2} \,\,\, .
\end{equation}
With this tuning of $V_0$, one finds
\begin{equation}
A^{\mu\nu|\rho\sigma} = - \frac{3}{D \, V_0} \left(\frac{\sum_i N_i \, (m_i^2)^{D/2}}{\sum_i N_i \, (m_i^2)^{D/2-1} }\right)
\left[ \left(\frac{D}{2}-1\right)(\eta^{\nu\rho} \eta^{\mu\sigma} + \eta^{\nu\sigma} \eta^{\mu\rho}) - \eta^{\mu\nu} \eta^{\rho\sigma} \right]  \frac{1}{q^2} + \cdots
\end{equation}
from which one infers that the Planck mass is given by
\begin{equation}
M_\Pl = \left[ \frac{\Gamma(1-D/2)}{6 \, (4\pi)^{D/2}} \sum_i N_i \, (m_i^2)^{D/2-1}\right]^{\frac{1}{D-2}} \,\,\, .
\label{eq:mp2}
\end{equation}
This reduces to the result in the theory with a single set of $N$ scalars in the limit that one or the other of the terms in the
sum dominates (or to a theory of $2N$ scalars of mass $m$, when $m_1=m_2=m$).  Note that for comparable $N_1$ and $N_2$, the Planck 
scale is set by the mass of heavier of the scalars, while the other can be much lighter.

In the two-scalar theory, the metric fluctuation that follows from the constraint of vanishing energy-momentum tensor can
be written
\begin{equation}
h_{\mu\nu} = \frac{D/2-1}{V_0} \left[\sum_{a=1}^{N_1} \partial_\mu \phi_1^a \partial_\nu \phi_1^a+\sum_{a=1}^{N_2} \partial_\mu \phi_2^a \partial_\nu \phi_2^a \right] - \eta_{\mu\nu}\,\frac{\Delta V}{V_0} + {\cal O}(1/V_0^2) \,\,\, ,
\end{equation}
which is algebraically equivalent to
\begin{equation}
h_{\mu\nu} = \frac{1}{V_0} {P^{\alpha\beta}}_{\mu\nu} ({\cal T}_1+ {\cal T}_2)_{\alpha\beta} + {\cal O}(1/V_0^2) \,\,\, .
\end{equation}
Following the same approach we used previously to write down the lowest-order effective interaction for a fundamental
graviton field, we find from Eq.~(\ref{eq:newLint}),
\begin{equation}
{\cal L}_{eff} = -\frac{1}{2} \, h_{\mu\nu} ({\cal T}_1 + {\cal T}_2)^{\mu\nu} \,\,\, ,
\end{equation}
where we have taken into account the two ways in which we may identify the fundamental graviton field.  This is of the expected
form, consistent with a universal coupling of the graviton to the two types of scalar fields in the theory.  The strength of the 
gravitational interaction is set by $M_\Pl$ in Eq.~(\ref{eq:mp2}).  The results of this section extend trivially to more than two
sets of distinct scalar fields. 


\section{The expectation value of the emergent metric} \label{sec:dff}

In the current set-up, where the energy-momentum tensor has a vanishing vacuum expectation value (VEV),  the expectation value of the metric fluctuation 
\be
h^{(1)}_{\mu\nu}= \frac{1}{V_0}P_{\mu\nu|\alpha\beta} {\cal T}^{\alpha\beta}
\ee
is zero, and the metric is flat.

\begin{figure}[h]
  \begin{center}
    \includegraphics[width=.5\textwidth]{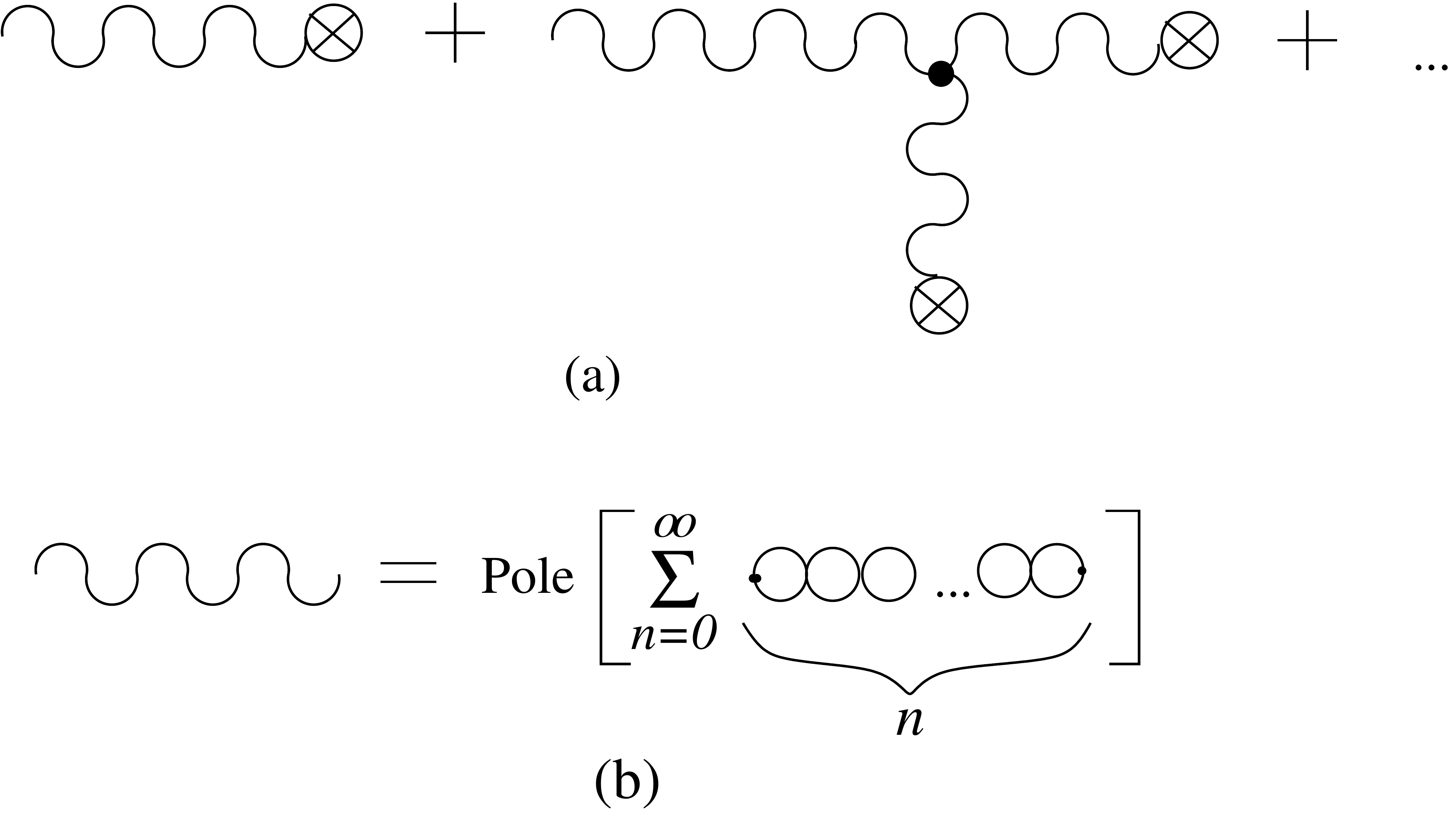}
    \caption{(a) The Feynman diagrams contributing to the VEV of the metric fluctuation in the presence of an external source.  The source is depicted by the crossed circle. The second
    diagram includes the three-graviton vertex. (b) Scalar loops that lead to the emergent graviton pole.}
        \label{fig:duff}
  \end{center}
\end{figure}

We would like to consider a scenario where a conserved source term is added to the action,
\be
\delta {\cal S}=-\frac 12 \int d^D x \, J^{\alpha\beta}(x) P_{\alpha\beta|\mu\nu} {\cal T}^{\mu\nu}.
\ee
In the presence of the the external source, $J_{\alpha\beta}$, the metric fluctuation can acquire a non-zero VEV
\be
\langle h^{(1)}_{\mu\nu} \rangle_J = \frac{1}{Z_J} \int {\cal D}\phi\, h^{(1)}_{\mu\nu} \, e^{i({\cal S}[\phi]+\delta {\cal S})} \, ,
\ee
where the normalization factor $Z_J$ is the partition function in the presence of the source. Diagramatically, the VEV is given by an infinite sum of which the lowest order terms 
are depicted in Fig.~\ref{fig:duff}.

This procedure mirrors at each step the general relativity calculation performed by Duff~\cite{duff} who showed that by summing such quantum tree diagrams, where the 
wavy lines in Fig.~\ref{fig:duff} denote the flat space graviton propagator, one can recover, for example, an order-by-order (in terms of mass) expansion of the Schwarzschild 
metric, if the source is taken to be spherically symmetric. By analogy, we expect the same result to hold in our theory, so that we obtain an emergent curved 
metric\footnote{For a different approach to incorporating curved backgrounds in the present framework, and which involves a modification of the starting point in 
Ref.~\cite{Carone:2016tup}, see Ref.~\cite{Chaurasia:2017ufl}.}. 

\section{Conclusions} \label{sec:conc}

In this paper, we have considered graviton interactions in a theory of emergent gravity proposed by  
Carone, Erlich and Vaman~\cite{Carone:2016tup}.   The original proposal was a theory of $D+N$ scalar
fields and was defined via a functional integral, subject to the constraint of vanishing total energy-momentum 
tensor.  The set of $D$ fields, called clock-and-ruler fields, had profiles that determined the 
background metric, but otherwise could be gauged away.  The composite graviton state was found to couple
to the non-vanishing energy-momentum tensor of the remaining $N$ scalar fields.   While the original work in
Ref.~\cite{Carone:2016tup} demonstrated the existence of a massless, spin-$2$ pole in scalar two-into-two 
scattering amplitudes in the large $N$ limit, the coupling of the graviton to itself, the universality of the coupling 
when additional scalars of differing mass are present, and an approach to incorporating curved backgrounds 
were not discussed.  Those issue were clarified in the present work.

In particular, we showed here that the effective three-graviton coupling inferred from scattering amplitudes involving
six external scalar lines is consistent with the expectations of the weak-field limit of general relativity.   The internal
graviton lines in the theory with a fundamental graviton field are contained within sums over loop sub-diagrams that each 
generate an emergent graviton pole.  The scattering amplitude arises by connecting these loop chains using the available four- and 
six-scalar vertices in the theory. This is similar to the way that non-Abelian gauge boson vertices were generated in the composite 
models of Ref.~\cite{suzuki}.  We showed by direct calculation that the Planck scale inferred from the induced three-graviton 
vertex, a function of the generally covariant physical regulator of the theory, agrees with the result found in the scattering calculation 
of Ref.~\cite{Carone:2016tup}.  Moreover, we explained in the present work why the perturbative expansion about a fixed 
background does not lead to an explicit breaking of general covariance, once loop corrections are taken into account.

We then considered extending the theory to the case where we begin with two sets of non-interacting 
scalar fields of differing masses.  This is a natural consistency check on the theory, since it would be 
unacceptable if, for example, there were a different effective Planck mass for each set of scalars.  We 
verified that the composite graviton couples to the energy-momentum tensor of each set of scalars with
a strength set by a single Planck mass, which is a function of the number and mass of each type of scalar field.  
We noted that in the case where there is a hierarchy in the scalar masses (assuming
the numbers of each type are comparable), the Planck scale is determined by the heavier scalar, while the other 
can be arbitrarily  lighter.  Finally, we briefly noted how to incorporate the physics of a curved background while still 
expanding about a flat one, by coupling the composite scalar operator that is identified with the metric to a 
classical source, implementing an approach known in the context of the weak-field limit of general relativity.

A number of issues are appropriate for future work.  In the type of scalar theories considered here, it is worthwhile
considering how the diagrammatic analysis generalizes in the case where scalar interactions are present in the
theory before the constraint of vanishing energy-momentum tensor is applied.  More precisely, how would the
analysis be modified if the scalar potential in Eq.~(\ref{c_2}) were to include, for example, a quartic term?   While the diagrams
included in the scattering calculation of Ref.~\cite{Carone:2016tup} are still present, additional diagrams that contribute to the
amplitude are expected; it needs to be shown that the graviton pole persists in this case.  Another interesting issue is
the generalization of the diagrammatic approach of Ref.~\cite{Carone:2016tup} to theories involving fields of spin-$1/2$ 
and $1$, required for applications to phenomenologically relevant quantum field theories.  Work in that direction is 
underway~\cite{ItNeverEnds}, and will be presented elsewhere.

\begin{acknowledgments}  
The authors are grateful for many fruitful discussions with Joshua Erlich.
The work of C.D.C. and T.V.B.C. was supported by the NSF under Grant PHY-1519644.
The work of D.V. was supported in part by DOE grant DE-SC0007894. D.V. would also like to acknowledge 
the hospitality of the College of William and Mary  Physics Department for the duration of this work.
\end{acknowledgments}

\appendix
\section{Three-graviton vertex decomposition} \label{sec:appendix}

The Feynman rule for the three-graviton vertex in the weak-field expansion of general relativity was used in Sec.~\ref{sec:3gmp}.
One can understand the Feynman rule from the following considerations:  The three-graviton part of $\sqrt{g} \, R$ can be decomposed
in terms of an operator basis, where each operator involves two derivatives and three graviton fields.  The following $14$ operators
form a suitable basis:
\begin{align}
O_1 & = h^2 \, \Box h \\
O_2 & = h\, \Box {h^\alpha}_\beta\, {h^\beta}_\alpha \\
O_3 & = h\, \partial_\mu {h^\alpha}_\beta \, \partial^\mu {h^\beta}_\alpha \\
O_4 & = {h^\alpha}_\beta \, \Box {h^\beta}_\mu \, {h^\mu}_\alpha \\
O_5 & = h_{\mu\nu} \, \partial^\mu h \, \partial^\nu h \\
O_6 & = h_{\mu\nu} \, h \, \partial^\mu \partial^\nu h \\
O_7 & = h_{\mu\nu} \, \partial^\mu {h^\alpha}_\beta \, \partial^\nu {h^\beta}_\alpha \\
O_8 & = h_{\mu\nu} \,  {h^\alpha}_\beta \, \partial^\mu \partial^\nu {h^\beta}_\alpha \\
O_9 &= h \, \partial^\mu h_{\mu\alpha} \, \partial^\nu {h_\nu}^\alpha \\
O_{10} &= h \, \partial^\mu \partial^\nu h_{\mu\alpha} \,  {h_\nu}^\alpha \\
O_{11} &= h \, \partial^\nu h_{\mu\alpha} \, \partial^\mu {h_\nu}^\alpha \\
O_{12} &= h^{\alpha\beta} \, \partial^\mu h_{\mu\alpha} \, \partial^\nu h_{\nu \beta} \\
O_{13} &= h^{\alpha\beta} \, \partial^\mu \partial^\nu h_{\mu\alpha} \,  h_{\nu\beta} \\
O_{14} &= h^{\alpha\beta} \, \partial^\nu h_{\mu\alpha} \, \partial^\mu h_{\nu\beta} \,\,.
\end{align}
The cubic terms of $\sqrt{g} R$ can then be written
\begin{align}
[\sqrt{g} R]^{(3)}& =\left[ \frac{1}{16} \,O_1 - \frac{1}{2}\, O_2 -\frac{3}{8}\, O_3 + \frac{1}{4} \,O_4\right] +\left[-\frac{1}{4}\, O_6 + \frac{1}{4} \,O_7 +\frac{1}{2} \,O_8 
\right] \nonumber \\
& + \left[ \frac{1}{2}\, O_{9} + O_{10} +\frac{1}{4} \,O_{11}  - O_{13} -\frac{1}{2}\, O_{14} \right] \label{eq:teq}
\end{align}
The momentum-space Feynman rule that follows from Eq.~(\ref{eq:teq}) leads to the quantity in parentheses in Eq.~(\ref{eq:3hmess}).   The prefactor 
of $4/M_\Pl$ arises from the following considerations: we start with the Lagrangian  ${\cal L}=-M_\Pl^2 \, [\sqrt{g}\, R] \,/ 2 + h_{\mu\nu} {\cal T}^{\mu\nu}/2$, and then 
rescale $h_{\mu\nu} \rightarrow 2 \, h_{\mu\nu}/ M_\Pl$ to place the graviton kinetic terms in canonical form.   This leads to the graviton-matter coupling 
$h_{\mu\nu} {\cal T}^{\mu\nu} / M_\Pl$, mentioned in Sec.~\ref{sec:3gmp}, as well as an overall factor of $-4/M_\Pl$ multiplying the Feynman rule that follows 
from Eq.~(\ref{eq:teq}).

\end{document}